\documentclass{article}

    \PassOptionsToPackage{numbers}{natbib}

\usepackage[final]{neurips_2022}
\usepackage[utf8]{inputenc} 
\usepackage[T1]{fontenc}    
\usepackage{hyperref}       
\usepackage{url}            
\usepackage{booktabs}       
\usepackage{amsfonts}       
\usepackage{nicefrac}       
\usepackage{microtype}      
\usepackage{xcolor}         
\usepackage{amsmath}
\usepackage{graphicx}
\graphicspath{ {./figures/} }

\newcommand{\vect}[1]{\boldsymbol{#1}}

\DeclareMathOperator*{\argmin}{arg\,min}

\title{Cross-Dataset Propensity Estimation for \\Debiasing Recommender Systems}

\author{%
  Fengyu Li \\
  Cornell University\\
  \texttt{fl334@cornell.edu} \\
  \And
  Sarah Dean \\
  Cornell University \\
  \texttt{sdean@cornell.edu} \\
}

\begin{document}

\maketitle

\begin{abstract}
  Datasets for training recommender systems are often subject to distribution shift induced by users' and recommenders' selection biases. In this paper, we study the impact of selection bias on datasets with different quantization. We then leverage two differently quantized datasets from different source distributions to mitigate distribution shift by applying the inverse probability scoring method from causal inference. Empirically, our approach gains significant performance improvement over single-dataset methods and alternative ways of combining two datasets. 
\end{abstract}

\section{Introduction}
Selection bias is one of the most prevalent sources of biases for recommender systems \cite{chen2020bias}. Selection bias happens when there is a pattern in the users' ratings that is unique to the training set. For example, in a recommender system for movies, users might mainly rate movies that are recommended to them, which is a small section of movies already tailored to the users' tastes \cite{pradel2012ranking}. However, the environment the recommender system is deployed on contains all candidates movies, which is not biased toward personal tastes. This discrepancy produces a misalignment between training and deployed settings, which is known as a distribution shift. Tackling the selection bias in a recommender dataset has been a constant challenge in designing recommender algorithms \cite{schnabel2016recommendations}.

When recommender systems are deployed in real-world platforms, clients are usually able to collect preference-associated data from different source distributions. These data, or feedbacks, are either implicit or explicit \cite{aggarwal2016recommender}, highly-quantized (e.g. binary) or less quantized (e.g. one to five stars). However, state-of-the-art recommender algorithms tend to focus on one dataset. We hypothesize that combing more than one datasets with different degrees of bias in the training procedure would make the model more robust against selection bias.

In this paper, we attempt to take advantage of datasets from differently biased sources with different levels of quantization. More specifically, we propose a way to feed both a highly quantized and a less quantized dataset to a gradient-based recommender algorithm, given the assumption that the highly quantized dataset contains less bias. Practically, this assumption stands true under many settings. 
Implicit datasets, which are usually more quantized, are often adopted because they are observed to be less biased towards personal taste \cite{amatriain2009like,jawaheer2010comparison}.

To ensure both datasets do not significantly lose their value in the presence of selection bias, we first examine the susceptibility to selection bias of differently quantized datasets from a source distribution. Our experiment shows that susceptibility to selection bias is not correlated with the way a dataset is quantized. 

Then, since a less-quantized dataset inherently contains more information than a more-quantized dataset \cite{widrow1996statistical}, we decide to use it as the training set of matrix factorization and use the more quantized dataset for deriving the inverse-probability-scoring (IPS) propensity estimator \cite{thompson2012sampling,imbens2015causal}, a causal inference approach applicable to matrix completion-based recommender algorithms. In this way, our cross-dataset learning framework empowers existing recommender algorithms by empowering highly quantized dataset. We carried out an experiment and found our method to outperform baselines by a significant margin.


\section{Related Work}
Prior works on overcoming selection bias-induced distribution shift via a propensity-based approach begins with the seminal paper \cite{schnabel2016recommendations}, which introduces the IPS method into recommender systems. Follow-up works aim at providing a learning-based or behavioral model of user feedback for propensity estimation \cite{joachims2017unbiased,yang2018unbiased}, which remains the central concern of this approach. While many works rely on a propensity matrix that is assumed to be known, e.g. \cite{something}, our work provides an idea of using a more quantized, implicit dataset for accurate propensity estimation.

The IPS method in our context is essentially a method of weighting training examples to correct the bias in the training data. Equivalent approaches such as importance weighting are widely used for domain adaptation in fields other than recommender systems \cite{sugiyama2007covariate,zhang2018importance}. 
Discussions on countering selection bias, e.g. in recommender systems, with IPS-based domain adaptation remain limited due to the hardness of evaluating propensities.

\section{Susceptibility to Selection Bias}
We first examine the susceptibility to selection bias of differently quantized data by manually introducing biased distributions of various degrees to the differently quantized training sets. It is crucial for us that the highly quantized dataset does not exhibit particular weakness when facing selection bias so that they can be properly adopted for propensity estimation.

\subsection{Simulated Environment for Controlling Bias and Quantization}
Since selection bias is uncontrollable in a dataset completely drawn from real-world, we have to adopt a simulated environment \cite{krauth2020offline} with both semi-synthetic and synthetic datasets explained below. In our environment, we propose the \textbf{softmax observation model} and introduce a hyperparameter $\beta$ to control the degree of bias. For a rating matrix $R$, the corresponding probability matrix of each rating being observed is $\Pr(R_{u,i}\text{ is observed}) = k\,\text{softmax}(\beta R_{u,i})$, where $k$ is set so that the expected proportion of observed ratings is controlled. In this observation model, a larger $\beta$ induces more significant bias at an exponential rate. In our experiment we assume constantly $10\%$ of ratings are observed among $1000$ users and $1000$ items. Quantization is simulated by mapping the same underlying continuous-valued rating matrix $R$ to $n$ different quantities evenly spaced between $0$ and $1$. In all experiments we used only 2, 3, and 5 as quantization levels.

\subsection{Datasets}
\textbf{Imputed ML100K Dataset.}
The ML100K dataset provides 100 thousand MNAR (missing-not-at-random) ratings across 1683 movies rated by 944 users and is a standard dataset used for recommender systems. Since we need the ground truth ratings for controlling bias, we use imputed ratings computed via standard matrix factorization \cite{krauth2020offline}. We normalize ratings between 0 and 1 for consistency with the other dataset. 

\textbf{Latent Factors Simulated Dataset.}
The latent factors dataset is a synthetic dataset that models users and item with Gaussian latent vectors to simulate preferences \cite{koren2008factorization,koren2009matrix}; both also have biases. The environment is provided by \cite{krauth2020offline}.

\subsection{Results}
\begin{figure}
  \centering
  \includegraphics[width=\textwidth]{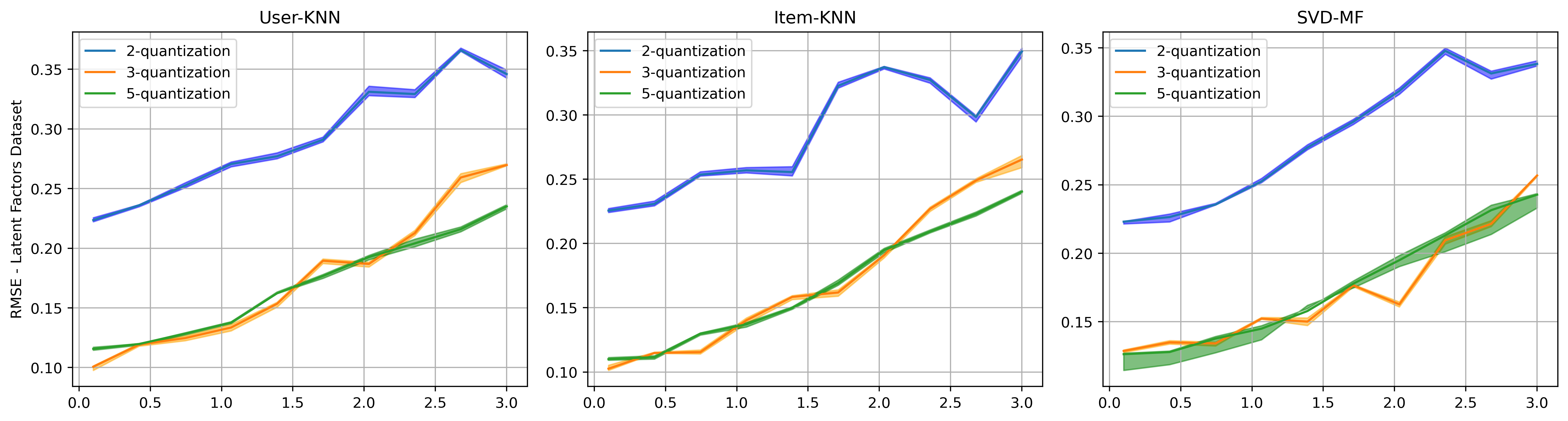}
  \caption{The effect of selection bias is not affected by quantization}
\end{figure}
Figure 1 shows the performance on the latent factors dataset of 3 classic algorithms: user-KNN, item-KNN, and SVD matrix factorization. Performance is evaluated by RMSE on the dense ground truth rating matrix. The three differently quantized variants come from the same source distribution (the dataset). Gaussian noise is introduced for simulating real-world user variability. Shaded regions denote the regions between quartiles overs 10 trials of re-sampled training data. 

Although the RMSE grows consistently as sampling bias $\beta$ increases, differently quantized datasets do not exhibit significantly different growth rates. The discrepancy between the RMSE of differently quantized datasets with the same $\beta$ can be ascribed to the inherent information loss from quantization \cite{widrow1996statistical}. We reason that since the effect of bias on rating predictions is consistent across quantizations, the subsequent effect on propensity estimation should be identical as accurate propensity estimation follows accurate rating predictions. Therefore, differently quantized datasets should also be equally susceptible to selection bias when used for propensity scoring. We thus conclude it is a viable approach to use the more quantized, less biased dataset for propensity estimation. Furthermore, Figure 1 visualizes the damaging effect of selection bias. 
The less quantized data with high sampling bias ($\beta=2.5$) performs just as poorly as the coarse binary dataset with no sampling bias ($\beta=0$),
suggesting the practical importance of debiasing recommender systems. 
The above experiments are repeated with the semi-synthetic ML100K environment and the conclusion holds.

\section{Cross-Dataset Propensity Estimation}
With the above result, we now turn to our cross-dataset matrix factorization model. Matrix factorization is a simple recommender model that decomposes the rating matrix based on known entries and then predicts the unknown entries \cite{koren2009matrix}. We integrate propensities as in \citep{schnabel2016recommendations} and formulated the recommendation problem as the empirical risk minimization framework below. 
\begin{equation} \argmin_{V,W,A} \frac{1}{N}\left(\frac{(Y_{u,i}- (V_u^TW_i + A))^2}{P_{u,i}} \right) + c \Vert A \Vert^2
\end{equation}
where $A = \{ b_u, b_i, \mu \}$ represents the standard bias parameters (offset), $Y$ is the ground truth ratings, $V$ and $W$ are the decomposed vectors, $\hat{Y} = V_u^TW_i + A$ is the predicted rating, $N$ is the number of observed ratings, and $c \Vert A \Vert^2$ is the regularizer. The inverse propensity scores $\nicefrac{1}{P_{u,i}}$ are multiplied to each rating during learning, which is analogous to re-weighting ratings based on their biases.

Denote the biased training set as $D$. we propose the \textbf{naive-bayes propensity estimator} (NBPE-MF) from a more quantized (binary) dataset $D'$. Essentially,
\begin{equation} 
  P_{u,i} = \Pr(Y_{u,i} \text{ is observed} \mid Y_{u,i} = r_{u,i}) = \frac{\Pr(Y_{u,i} = r_{u,i} \mid Y_{u,i} \text{ is observed})\Pr( Y_{u,i} \text{ is observed})}{\Pr(Y_{u,i} = r_{u,i})},
\end{equation}
where $r_{u,i}$ denotes the value of a rating. The numerator can be easily approximated from the dataset, but the denominator requires additional data from less biased sources. In this paper, we estimate the denominator from the less biased dataset using a categorical naive-bayes model. We use $D'$ as a mask to hide biased ratings in the training set from propensity estimation. In other words, $Y_{u,i} \text{ is considered observed}$ only when $D'_{u,i} \text{ is observed}$:
\begin{equation} 
  \Pr(Y_{u,i} = r_{u,i}) = \frac{\sum_1^N \vect 1_{(Y^D_{u,i} = r_{u,i})} \cdot \vect 1_{(Y^{D'}_{u,i} \text{ is observed})} }{N}.
\end{equation}
If the corresponding rating is not captured in $D'$, we set its propensity to the average propensity.

\section{Experiment}
We designed experiments to verify the performance of our cross-dataset model. We trained on two datasets (as in section 3) with $\beta =1$ in training set and used three baseline algorithms for comparison. In this paper we assumed the highly quantized dataset is unbiased with $\beta=0$, which is probable when the implicit dataset reflects users' spontaneous choices. Empirically, our framework will work as long as the highly quantized dataset is less biased. We selected the root mean square error (RMSE) and the mean absolute error (MAE) over the dense ground truth rating matrix as performance metrics. 

\subsection{Baselines}
\textbf{Matrix Factorization (MF.)}
As a simple baseline, we adopt the standard matrix factorization that does not adopt propensity estimation or importance weighting and uses only the less quantized training set.

\textbf{Naive Propensity Estimator (NPE-MF.)}
The naive propensity estimator naively estimates the propensity scores from the already biased training data and plugs in the results to equation 1.

\textbf{Mixing Datasets Matrix Factorization (MD-MF.)}
To analyze whether our cross-dataset model more efficiently exploits all existing data, we create a larger dataset by mixing the two differently quantized datasets and train the MF algorithm on it. To do this, we tried multiple schemes to upscale the more quantized dataset to be less quantized and found that this has minimal influence on the results.

\subsection{Results}
Table 1 shows the the experiment's results, each entry as the average of five independent trials. Averaged across datasets and metrics, our approach gains $23.1\%$ reduction in error compared to matrix factorization, $23.1\%$ compared to naive propensity estimator, and $22.1\%$ compared to mixing datasets matrix factorization. 

Additionally, we observe that NPE-MF and MF have very similar performance across all settings. This suggests that a IPS-based model will not gain any performance boost if propensities are inaccurately estimated. Furthermore, we argue that our model is data-efficient, i.e., it beats the competitor model (MD-MF), demonstrating that utilizing less biased dataset for propensity estimation is a more viable approach than mixing it with other data, though the latter is sometimes a common engineering practice. We nudge the way two datasets are mixed in MD-MF and discover similar results. On our benchmarks, mixing datasets MF consistently achieves less than $2\%$ advantage compared to using only one dataset.

We conclude that our cross-dataset model significantly outperforms baselines and provides a efficient way to make use of more quantized (implicit) data. 

\begin{table}
  \caption{Test set RMSE and MAE for NBPE-MF and baselines}
  \label{sample-table}
  \centering
  \begin{tabular}{lllll}
    \toprule               
                     & \multicolumn{2}{c}{ML100K} & \multicolumn{2}{c}{Latent Factors}                                     \\
    \cmidrule(r){2-5}
                     & RMSE                       & MAE                                & RMSE            & MAE             \\
    \midrule
    MF               & 0.1046                     & 0.0833                             & 0.1331          & 0.1045          \\ 
    NPE-MF           & 0.1048                     & 0.0834                             & 0.132           & 0.1049          \\ 
    MD-MF            & 0.1044                     & 0.083                              & 0.1299          & 0.1027          \\
    \textbf{NBPE-MF} & \textbf{0.076}             & \textbf{0.062}                     & \textbf{0.1065} & \textbf{0.0841} \\ 
    \bottomrule
  \end{tabular}
\end{table}

\section{Conclusion}
We propose an effective and data-efficient method of combining differently sourced datasets for training recommender systems. Our approach provides yet another way of accurate propensity estimation and explores a new potential of cross-dataset joint learning. We demonstrate that our method outperforms naive propensity estimation and straightforward dataset mixing. We believe real-world recommender systems will benefit from data-centric frameworks like ours.

\section*{Acknowledgements}
This work is supported in part by a Cornell CSURP Grant and by a research gift from Wayfair.

\bibliography{paper}


\end{document}